# An Object Web Seminar: A Retrospective on a Technical Dialogue Still Reverberating

James J. Cusick
*Dept. of Electronic and Computer Engineering*
*Ritsumeikan University*
Kusatsu, Shiga, Japan
j.cusick@computer.org

***Abstract***— *Technology change happens quickly such that new trends tend to crowd out the focus on what was new just yesterday. In this paper the peak popularity of the confluence of Object Technologies with early Web adoption is explored through the content of a seminar held in 1999. Distributed architectures were undergoing significant change at this point, and deeper software capabilities were just beginning to be broadly accessible over the Internet. The Object Web arose and was infused with new development tools reflecting these capabilities and allowing design of applications for deployment during the early days of the World Wide Web. This conference discussed the history, evolution, and use of these tools, architectures, and their future possibilities. The continued dominance of these approaches although under different names is demonstrated even though the term Object Web has receded in use. Favored newer offerings such as Kubernetes and microservices still model the core design attributes of the Object Web for example. Aside from connecting this seminar to relevance in the software world of today this paper also touches on the early AI tools demonstrated in this seminar a quarter century ago and how the popularity wave of any given technology might affect the current focus on AI technology offerings.*

## I. Introduction

Technological innovation often appears to move more rapidly than we can track. With so many things positioned as new breakthroughs it can be hard to remember which ones developed more slowly and have been with us for a long period of time or evolved from earlier work. This fact jumped out at me recently upon finding a document from 1999 [1]. This document provided an introduction for a seminar on the Object Web held at AT&T that year. One may assume that very little discussed at that seminar would be relevant in today's world. Surprisingly, upon review, this particular artifact demonstrated that this was not the case. Instead, nearly all the technologies discussed at that event (with one notable exception) have continued to evolve, provided the foundations for today's software architecture, and remain in active development.

In this current paper, the background of this seminar, its agenda, contributors, and the details of its content will be explored. Of particular interest is the historical treatment of the development of the Object Web as presented in the seminar introduction which traces the origins of many popular technologies of today such as Python, distributed computing, and the cloud. Furthermore, those technologies which had less success (especially CORBA) will also be put into context. Finally, the seminar also touched on AI technologies of the day. A linkage will be presented between the fortunes of the Object Web and the intense focus on burgeoning AI tools and applications today.

## II. Background

The Internet found mass appeal through the invention of the World Wide Web or what would simply be called the Web. Even after a major financial setback in 2001 referred to as the dotcom crash, things continued on with even greater success [2]. Powering this transformation were a variety of specific technologies including HTML, broadband, and scalable core network switches. Each of these technologies (and more) remain prevalent today, however, one technology which at the time was seen as the next great domain for innovation has faded from conversation but not from usage.

The Object Web was a major focus during the late 1990s [3]. Many technologists saw this convergence as the next dominant technological environment. Instead, something different happened. The focus on the Object Web and the term itself seemed to recede from view while simultaneously the capabilities of the Object Web and the predictions around it came into quiet dominance. Today, the Object Web surrounds developers and technology consumers in nearly equivalent ways as originally imagined and forecast years earlier, but without any fanfare or popular acknowledgement in the way that AI is currently enjoying a moment in the public imagination.

In fact the creator of the World Wide Web and its key protocols spoke to the underlying reliance on objects in powering the Web [4]:

> *"The Web may be viewed as a collection of objects and indeed, as originally designed, had an extensible space of methods to be applied to those objects." – Bernes-Lee*

At the time, Object Technology including Object-Oriented Programming languages, design methods, data management, and more, dominated the software development landscape. In parallel and at roughly the same time, the Internet and the Web began to dominate not only the technical landscape but the business environment and large parts of the economy as well. This generated significant interest in the interplay of these technologies and how they would develop together going forward. As a result, the term Object Web came into common use in the industry in the mid to late 1990s and remained popular for a number of years afterwards.

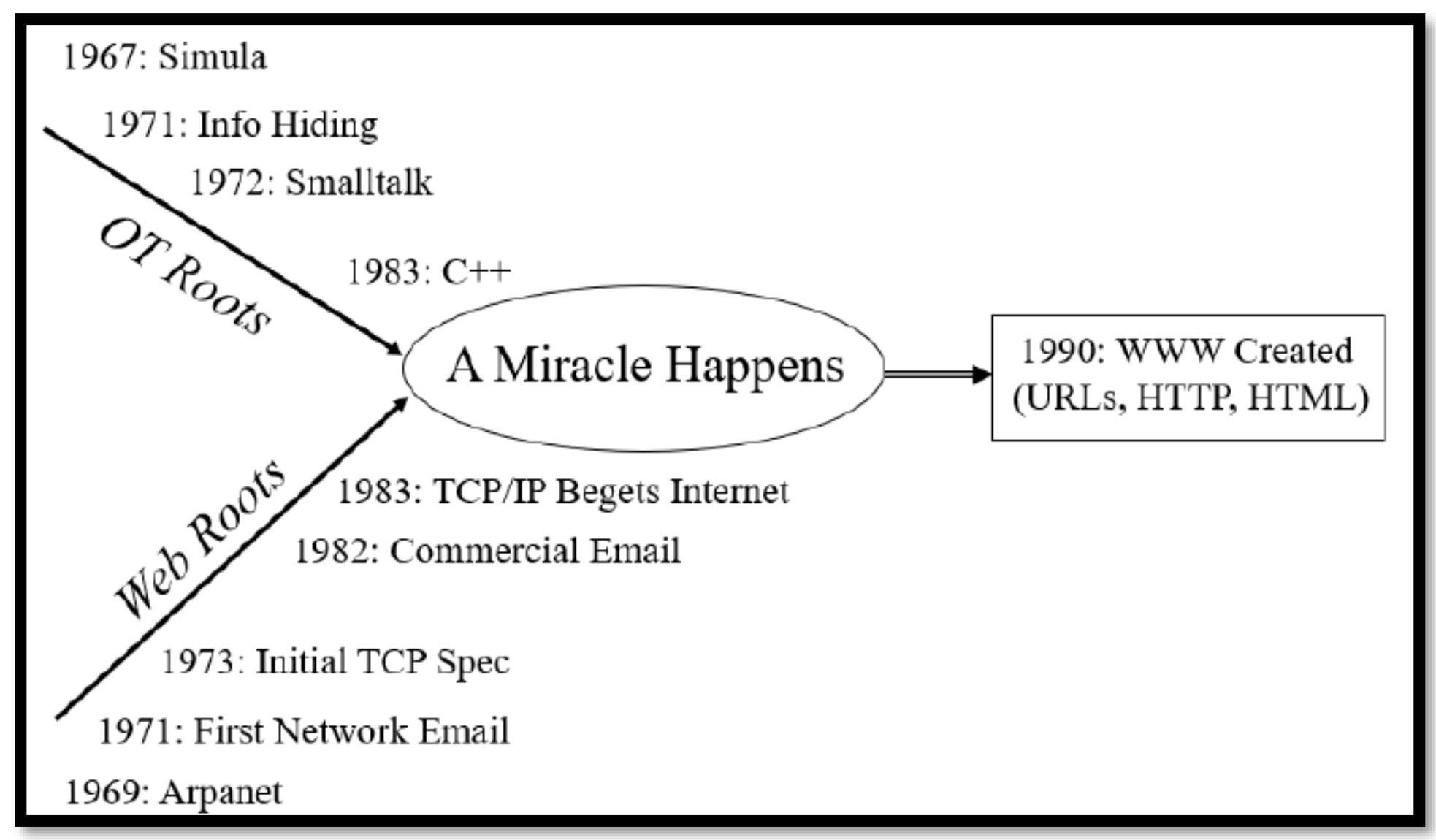

*Figure 2 – Initial diagram from Object Web Seminar introduction depicting the origins of both Object Technology and of Internet Technologies.*

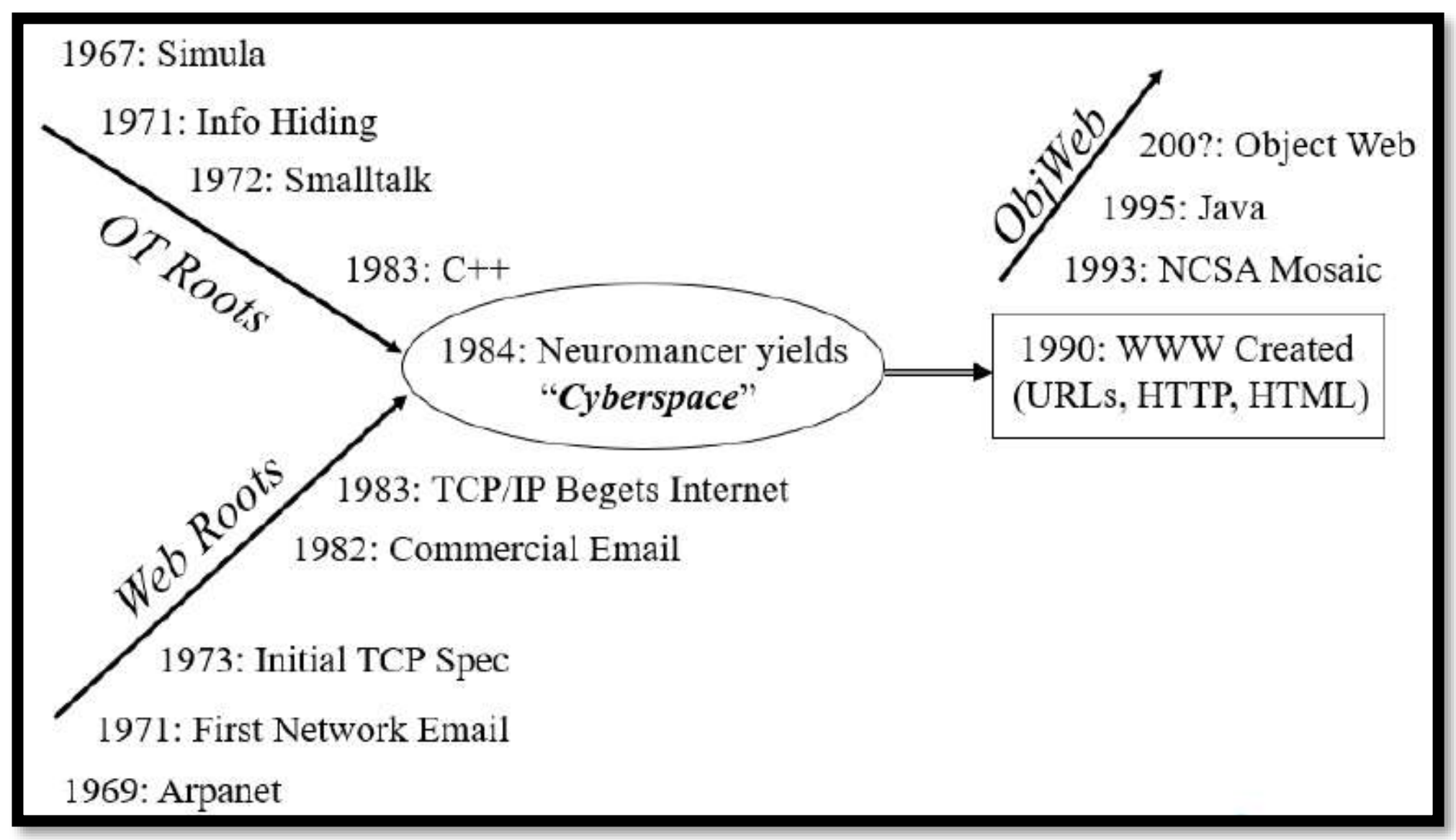

*Figure 1 – Seconfdary diagram from Object Web Seminar depicting the projected growth of the Object Web.*

## III. Object Web Seminar Description

To explore the topic of the Object Web a seminar was organized and held. This seminar entitled "Objects on the Web" was one of a series of events organized within AT&T Labs to educate, promote, and foster debate around technical trends and capabilities. The seminar was offered by AT&T Labs' SysLab (System Innovation Lab) whose mission was to bring together developers, managers, researchers, and vendors to explore new developments in technology. Over 100 managers and engineers attended the seminar from AT&T and AT&T Labs as well as partner organizations.

Seminar presenters came from AT&T Labs, IBM, and WebLogic which pioneered the term "application server" through its Java J2EE offering. Additionally, a group of vendors demonstrated software and other products. These vendors included IBM with WebSphere, Sun Microsystems, BEA, Object Design (later eXcelon) which developed ObjectStore an early and successful Object database, and a &T Labs Research with a tool called DajaVu. This tool was developed by the AI Research team led by Yann Le Cunn [5]. Le Cunn worked at Bell Labs and AT&T Labs from 1988 to 2002 including at the time this seminar was held although he was not in attendance. However, the DejaVu tool, an offshoot of his lab's AI document object research, provided an open solution to document compression at a time when PDF was a closed technology as was demonstrated on that day. The remainder of the event consisted of a keynote, several detailed technical talks, and finished with a set of parallel tutorials.

Specific technologies discussed during the seminar's presentations included Object Technologies in general, CORBA (Common Object Request Broker Architecture), Enterprise Java (EJB-JavaBeans), WebLogic's J2EE server, wrapping legacy applications with object interfaces, and hands-on tutorials in Python, EJB, and CORBA performance. Interestingly the Python tutorial was advocated by one of the development visionaries of AT&T Labs who began working with Python in the mid-1990s and saw it as a strong future technology candidate which in today's environment has proven to be the case.

## IV. Tracing the Precursors of the Object Web

The seminar's introduction presentation focused on a definition of the Object Web, the precursor technologies leading to it, and where the Object Web would be going in the future. The historical summary of two major technologies, objects and the web, up to that time was outlined. This history, while brief, remains accurate. This paper enhances the original historical notes provided in the introduction during the seminar itself by adding specific supporting references below. However, what makes this seminar content interesting is both what it got right about the future and what it got wrong.

To explain where the Object Web came from and where it was going, the introductory presentation divided the terms and layout of their ancestry and heritage. In Figure 1 the initial aspects of the legs of this milieu were presented: 1) the roots of Object Technology, and 2) the roots of the Internet and the Web. Object Oriented development itself from analysis to design to programming saw major adoption in the 1990s. Formally, an object is considered an abstraction from a given problem domain reflecting system capabilities and information encapsulating attributes and services [6].

A reasonable starting point for the history of Object Technology realistically begins with Simula [7]. Shortly thereafter, Parnas [8], formalized a key concept used in Object Technology which defines modularization and information hiding. With the emergence of Smalltalk [9] and then C++ from Stoustrup [10] powerful language options became prevalent allowing for the broad adoption of object-oriented designs and program construction.

On the other leg of the diagram the evolution of the Internet and the Web are summarized. Foreshadowed by the visionary paper from Bush [11], which predicts a universal computing environment with hypertext and two-way video, such an environment eventually began taking shape. First, we had the introduction of ARPANET [12], then the development of email by Tomlinson [13], and importantly, the specification of TCP/IP [14] providing for a truly interoperable and distributed computational model. These capabilities were then stitched together in the fictional novel Neuromancer [15] where the term Cyberspace was established and subsequently became both a part of the popular imagination as well as a metaphor reflecting the global computing domain then coming into form.

With this underlying infrastructure in place, The World Wide Web was built on top of these tools and ideas [16]. The early browsers such as Mozilla and NCSA [17] opened up access to Internet based resources for the general public. The move from universal addressing of static pages to the realization of dynamic web content was bolstered by the introduction of Java [18] and backend connectivity using CGI and RMI protocols. This general approach became known as distributed object technology based systems [19, 20].

## V. Other Seminar Topics

Following the introduction to the seminar, a keynote was presented followed by several technical talks, and a few hands-on tutorials [21]. The keynote was provided by the late Prof. William Tepfenhart of Monmouth University and formerly of Bell Laboratories. Prof. Tepfenhart gave an overview and short history of both the Web and the emergence of various object models used on the Internet, and then proceeded to describe how the Web, by its nature, lends itself to an object model. Prof. Tephenhart explained that the future of objects on the web was with CORBA. He did not view EJB as a competitor to CORBA, but a complement.

The WebXpress team focused on server side processing approaches. They explained that Enterprises were finding numerous reasons to employ a server side model for business logic (for example, maintainability, scalability, security, integration). Furthermore, Java was seen as the emerging language of choice for network enabled applications. The detailed presentations were rounded out by discussions on EJB "Container Services", legacy wrapping approaches, CORBA performance topics, and the hands-on tutorials covering Python and distributed computing models.

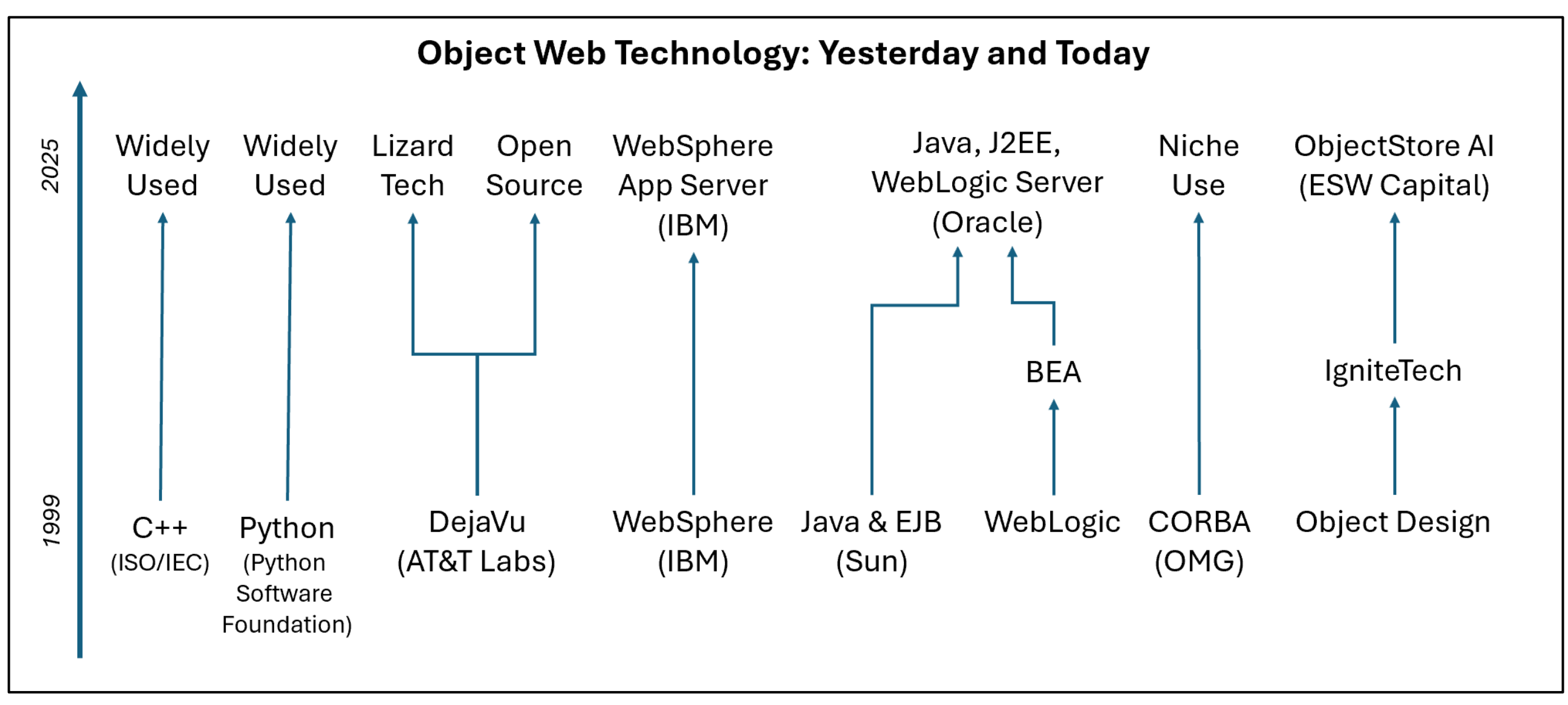

*Figure 3 – A 25-year notional history of technology and vendor evolution of products discussed in the Object Web.*

## VI. Object Web Technical Legacy

While reviewing the Object Web, its underlying technologies, and their histories may be interesting, the subject becomes even more interesting when connecting to the technical legacies evident some 25 years later. Scratching the surface of nearly any software technology today quickly unearths the very tools, languages, architecture patterns, and even vendors who were front and center a quarter century ago. This represents an especially long period of time in technology circles. For an industry famously filled with short-lived companies, it is almost unheard of that those technologies and vendors highlighted during this one seminar 25 years ago continue to exist today and in some cases provide highly relevant and leading technical solutions or at minimum design approaches such as object interfaces to cloud based containerized services.

### A. *Vendor Evolution*

Limiting the discussion to only those companies highlighted in the Object Web seminar we see that both AT&T Labs and IBM remain. For nearly all the other companies involved some change did indeed come but not as dramatically as might be predicted. The heritage and evolution of the companies involved in the seminar is depicted in Figure 3. The programming languages of C++ and Python remain widely in use and have largely increased in adoption [22]. For the vendors, WebLogic was bought by BEA and then BEA was bought by Oracle. Sun of course was also acquired by Oracle. Object Design eventually became a part of IgniteTech's software catalogue and is currently marketed as ObjectStore AI. Progress Software sold Object Design to venture capital company ESW Capital.

As a result, all of these businesses continue to this day in one form or another which is somewhat surprising. Most took their distributed and object-oriented technologies to scale during the expansion of the Internet age and are now continuing to build on these platforms, often relabeling them as AI tools or services such as with ObjectStore. Such "AI washing" or labeling pre-existing technologies as AI is common today and ironically follows the same trend in the 1990s when many earlier classic tools were painted with an OO brush to maintain market relevancy.

Naturally, to stay relevant, these companies have needed to evolve their products and technologies. While the essence of an object-oriented distributed computing model has been retained by these products, most solutions have moved away from architectures using RMI or ORBs with IIOP access. Instead, modern applet and servlet architectures proliferate but are at their heart remote objects invoked via a defined API call such as the Java servlet class. In the case of WebLogic their product, formerly called EAI for Enterprise Application Integration server, is now the Oracle WebLogic Integration (WLI) platform. The latest trend is providing autonomous logic within such objects (which reside on the web) and labelling them as "Agentic AI" [23]. Then each object or service is accessed across the wire (in a distributed manner) via an interface wrapper defined as an object. This closely resembles the core architectures of the distributed Object Web of 1999 with the addition of some level of programmatic autonomy.

### B. *Architecture Evolution*

Based on the history outlined above, it appears that within the distributed object web software ecosystem, continuity has proven to be more common than revolution. Early web solutions relied on static content using straightforward hypertext pages. Soon, client-side architectures with HTML, CSS, and JQuery dominated in the 1990s. This was expanded with Server Side MVC application architectures using HTML5 and Stateful Session management. Eventually RESTful Services End Point integration with WebAPI (Business Logic and Data Access Layer) approaches proliferated in the 2000s. In many cases legacy systems were "webified" by exposing the mostly unchanged software asset or system to the web using standard internet protocols.

This evolution has continued to today's cloud environment where leading compute environments such as Azure Services offering host functionality. This offering provides Kubernetes containers and orchestration platforms which mirror the encapsulation and remote invocation of objects [24]. In fact, whether using REST APIs or OCI (Open Container Initiative) compliant objects, such instances are in fact simply loosely coupled web-based objects echoing the early Object Web designs.

In the case of AWS both Python and Java are supported and backed by the Amazon S3 object storage capability. More broadly, Microservices have become the common design choice for most applications which cement the object-oriented principles of domain abstraction, encapsulation, and defined interfaces in place [25]. Even popular web UIs today are built with object technologies like React.js [26]. React uses a virtual DOM (Document Object Model) which traces its lineage to the mid-1990s and Netscape's browser as well as the W3C version to provide performant front-end rendering. Again the concept of the DOM was born at the time when the Object Web was the paramount computing model.

### C. *Objects Everywhere*

Perhaps one contributing factor to the lack of focus on the Object Web in recent years is the success of the technology and its diffusion throughout the many child technologies inheriting its characteristics. From the OOP languages already mentioned such as C++ and Python, most new languages contain object-oriented features. From the commonly used DOMs and BOMs (Browser Object Models) to architecture patterns including applets, servlets, and microservices, objects are commonplace. In fact, it would be difficult to identify current tools, applications, or services on the Internet which do not use some form of object technology. In away the anonymity of the Object Web stems from its ubiquitous and strong success.

A sampling of just a few current technologies melding the web, objects, and other more recent technologies include the following items:

1. Amazon's S3 is an object storage environment which is object-oriented and operates over the web (or cloud). This is a simple and ubiquitous example of how the Object Web has continued and evolved in use to today's computing world.
2. The HTML <object> element itself which represents an external resource, such as an

image, a nested browsing context, or a resource to be handled by a plugin [27].

3. The emerging applications of the spatial web with objects [28]. In this use case nearly all IoT devices are seen as linked together an providing a rich mesh of data to the user via an object-oriented array of application nodes.

## VII. Parallels to Today's AI Boom

In looking back at this seminar on the Object Web as a snapshot of where software development trends were moving at the end of the last century, the parallels to the ongoing boom in AI technology takes on a different contour. As we can see from the discussions during the seminar, some of the thinking about the future direction of certain components within the current software landscape was inaccurate, which should not be surprising. However, what is more surprising is how the preponderance of the analysis and technical foundations on display at the seminar remain valid today. In fact, many technical examples have only expanded and grown more capable and underpin nearly all software applications. This is true even for some mainframe environments but may be much more limited in specialty domains such as some device control or embedded applications.

When we cast this observation onto the current flourishing of AI related tools, technologies, and applications [29] we can draw a few general conclusions. First, the earliest AI concepts and technologies were developed in the 1950s and 1960s with a significant level of investment and innovation centered in the 1980s [30]. During this earlier period the focus was on expert systems, rule-based systems, and later machine learning as well as robotics and computer vision. Many breakthroughs were witnessed at this time prior to the "AI Winter" where investment was curtailed due to complexity barriers and other factors.

Nevertheless, AT&T, where the Object Web seminar was organized and held, never stopped its work in AI. In fact many Bell Labs software advances underpin both current computing models and many AI advances and tools [31]. Moreover, as mentioned above, one of the company's AI related tools (DejaVu) made an appearance at the Object Web seminar. In earlier work dozens of software development tools were evaluated and codified as company-wide standards [32]. Since the company's applications had special needs in the Artificial Intelligence arena several were included in the standards. They were developed by the AI Research team within AT&T Bell Labs or with external partners. Specifically, the tools were C5, CLASSIC, MLT, PLISP, OD, and R++ [33]. In Figure 4 an artifact from those days (a company provided coffee mug) lists many of these tools confirming their names. Unfortunately, documentation on these tools today is scarce. However, they did exist as the mug attests to. Which is a key part of the story here.

In essence, what we see with both the earlier boom in Object Web technology and this earlier phase of AI development is that some solutions will prevail and others will fade away. Naturally no one had a crystal ball back in 1999. Similarly no one has one today either. Yet it is a fair bet that a good portion of the AI technologies in play now will follow the same cycle of development, success or failure, acquisition or growth, and finally, absorption into the general fabric of the software and technology infrastructure supporting society. Of note is that knowledge classification tools like C5 and rule based programming tools like R++ represent fundamental AI technologies which are being used by today's AI toolsets [34]. Just like the distributed computing models of the 1990s led to today's containers, microservices, and remote object invocation APIs, the AI tools and languages of the 1990s seen from Bell Labs and other groups of the time helped set the foundations for the "overnight" success of companies like OpenAI.

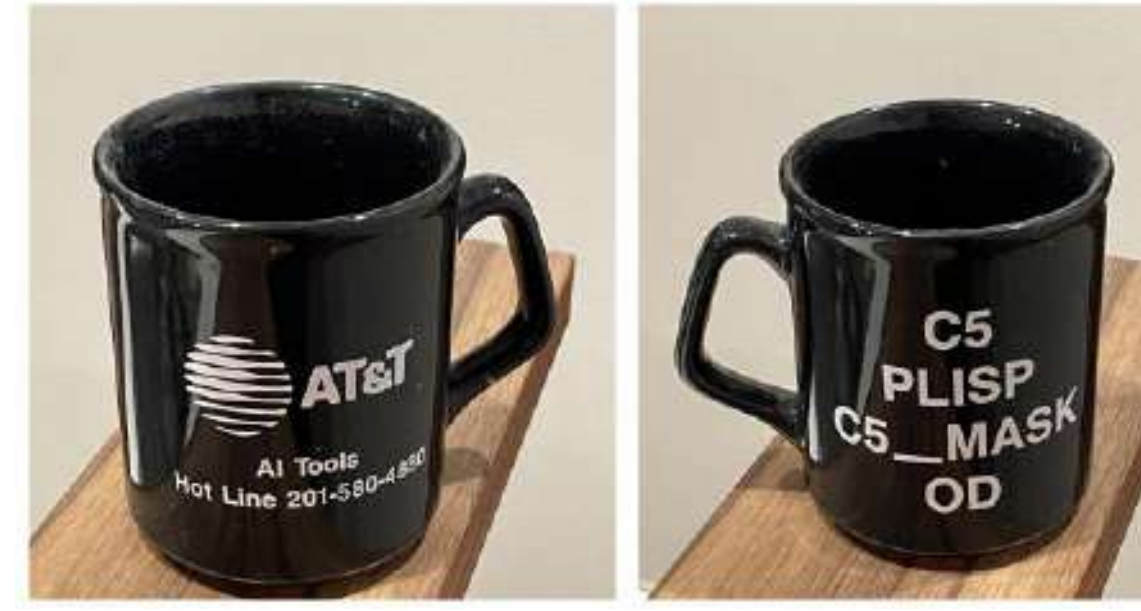

*Figure 4 – A souvineir from the AT&T Bell Labs AI Tools Research Group in 1996 still around in 2026..*

## VIII. Conclusions

As we have seen, the Object Web seminar held many years ago in 1999 as organized by AT&T Labs was both topical and prophetic if not perfectly so. The introduction to the seminar provided a solid history and evolution of both Object Technologies and the Internet and World Wide Web. In addition to this background a look down the road was also provided predicting wider adoption of these technologies. This did indeed happen over the ensuing years with the primary exception of the very limited long term use of CORBA. However, all around us today we have object-oriented programming languages, data stores, distributed architectures with remove invocation of objects through defined methods per class or object, and a variety of packaging and operational abstractions in various cloud environments. It seems this seminar predictions played well.

Additionally, as a case in point, the vendors and their technologies present for the seminar also seemed to weather the years quite well. Nearly all the companies taking part in the seminar remain in business as part of a different organization. Thus, aside from the observation that the Object Web in many aspects continues to dominate our application environment today, we also see that many software companies can build a lengthy track record in the industry.

Finally, this technology example provides some insight into understanding the current boom and/or hype surrounding AI. During the seminar back in 1999 some AI tools were presented and still others were part of the AT&T standards via the research department's efforts. While most of these tools have not made it forward to today's marketplace and the current boom in AI technology, the work done in those days did provide for some foundational underpinnings of the tools popular today. Based on this experience it might be fair to say that like the Object Web, which is all around us now but under different labels, perhaps the AI tools we see today will also be with us for years to come but there will be vendor consolidation, technical standardization, and less of a winner take all scenario. That is, if the history of the Object Web is any guide.